\documentclass[aps,prl,twocolumn,superscriptaddress,showpacs]{revtex4}
\usepackage{amsmath,amsfonts,bm,graphicx,amssymb,rotate}

\begin{document}

\title{Graphene Antidot Lattices -- Designed Defects and Spin Qubits}

\author{Thomas~G.~Pedersen}
\affiliation{Department of Physics and Nanotechnology, Aalborg
University, DK-9220 Aalborg East, Denmark}
\author{Christian Flindt}
\affiliation{MIC -- Department of Micro and Nanotechnology, NanoDTU,
Technical University of Denmark, DK-2800 Kgs.\ Lyngby, Denmark}
\affiliation{Laboratory of Physics, Helsinki University of
Technology, P.\ O.\ Box 1100, 02015 HUT, Finland}
\author{Jesper~Pedersen}
\affiliation{MIC -- Department of Micro and Nanotechnology, NanoDTU,
Technical University of Denmark, DK-2800 Kgs.\ Lyngby, Denmark}
\author{Niels~Asger~Mortensen}
\affiliation{MIC -- Department of Micro and Nanotechnology, NanoDTU,
Technical University of Denmark, DK-2800 Kgs.\ Lyngby, Denmark}
\author{Antti-Pekka~Jauho}
\affiliation{MIC -- Department of Micro and Nanotechnology, NanoDTU,
Technical University of Denmark, DK-2800 Kgs.\ Lyngby, Denmark}
\affiliation{Laboratory of Physics, Helsinki University of
Technology, P.\ O.\ Box 1100, 02015 HUT, Finland}
\author{Kjeld~Pedersen}
\affiliation{Department of Physics and Nanotechnology, Aalborg
University, DK-9220 Aalborg East, Denmark}

\date{\today}

\begin{abstract}
Antidot lattices, defined on a two-dimensional electron gas at a
semiconductor heterostructure, are a well-studied class of man-made
structures with intriguing physical properties.  We point out that a
closely related system, graphene sheets with regularly spaced holes
(``antidots"), should display similar phenomenology, but within a
much more favorable energy scale, a consequence of the Dirac fermion
nature of the states around the Fermi level. Further, by leaving out
some of the holes one can create defect states, or pairs of coupled
defect states, which can function as hosts for electron spin qubits.
We present a detailed study of the energetics of periodic graphene
antidot lattices, analyze the level structure of a single defect,
calculate the exchange coupling between a pair of spin qubits, and
identify possible avenues for further developments.
\end{abstract}



\maketitle

Graphene is the rapidly rising star of low-dimensional materials.
Following the initial reports on fabrication by mechanical
peeling~\cite{Novo2004} and epitaxial growth~\cite{Berg2004}, this
exceptional material has stimulated considerable
experimental~\cite{Geim2007} and theoretical
research~\cite{Kats2006} as well as proposals for novel electronic
devices~\cite{Ryce2007}. The promising prospects for graphene
devices are based on several remarkable properties. Mainly, the
sample quality and mobility (exceeding 15000
cm$^2$/Vs~\cite{Geim2007}) can be very high. In addition, patterning
of such monolayer films by e-beam
lithography~\cite{Geim2007,Berg2006} with features as small as 10
nm~\cite{Geim2007,Han2007} is possible. Very recently, spintronics
devices have been considered~\cite{Hill2006}. The incentive for
graphene based spintronics lies partly in the long spin coherence
time that is characteristic of carbon-based materials. This also has
obvious advantages within the field of solid-state quantum
information processing, where confined electron spins have been
promoted as carriers of quantum information~\cite{Loss1998}. Being a
light element, carbon has a rather small spin-orbit coupling, and,
moreover, the predominant $^{12}$C isotope has a vanishing hyperfine
interaction. This makes graphene, at least in principle, a superior
material compared to existing quantum computing implementations in
GaAs~\cite{Trau2007,Pett2005}.

Antidot lattices, defined on semiconductor heterostructures, display
many intricate transport properties, in particular in magnetic
fields where the competing length scales lead to rich
physics~\cite{Weis1991}. In this Letter we wish to draw attention to
the possibility of forming antidot lattices on graphene.  As
mentioned above, state-of-the-art e-beam lithography has been used
to carve graphene nanoribbons with feature sizes down to tens of
nanometers. We propose to use similar techniques to create regular
holes in the graphene sheet, in order to form antidot lattices. The
antidot lattice has the important consequence that it turns the
semi-metallic graphene into a gapped semiconductor, where the size
of the gap can be tuned via the antidot lattice parameters. As our
analysis shall show, this electronic structure can be manipulated
further so as to create coupled electron spin qubits, thus
suggesting that these perforated graphene sheets are a promising
platform for a large-scale spin qubit architecture. Localized spin
qubit states can be formed in the antidot lattice by deliberately
omitting some of the antidots. This idea has previously been
analyzed for the two-dimensional electron gas in e.g.\ GaAs
heterostructures~\cite{Flin2005}. As we will now argue, moving to
graphene has three major advantages: (i) increased coherence time;
(ii) favorable energy scale of the defect states; and (iii)
increased lateral confinement.

\begin{figure}
\includegraphics[width=0.44\textwidth]{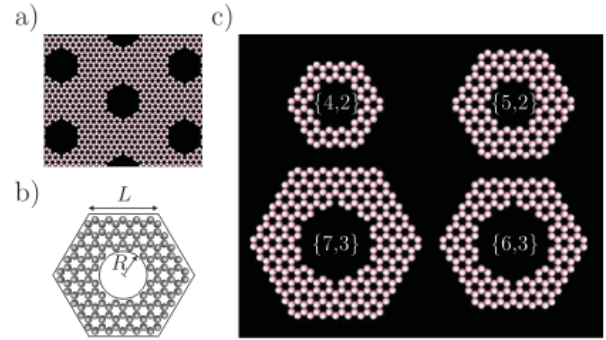}
\caption{Illustration of the triangular antidot lattice (a) with a
unit cell characterized by side length $L$ and hole radius $R$ (b).
In (c), several examples with corresponding $\{L,R\}$ parameters are
shown. }\label{fig1}
\end{figure}

The proposed antidot lattice is simply a triangular array of holes
in a graphene sheet, as illustrated in Fig.\ \ref{fig1}a. The
lattice consists of hexagonal unit cells as shown in Fig.\
\ref{fig1}b, in which a roughly circular hole is created. We
characterize the structure by the side length $L$ of the hexagonal
unit cell and the radius $R$ of the hole, both measured in units of
the graphene lattice constant $a\approx 2.46$ {\AA}. A lattice is
designated by the notation $\{L,R \}$. Note that while $L$ is an
integer, $R$ can be non-integer. As is evident from the examples in
Fig.\ \ref{fig1}c, $L$ is equal to the number of carbon atoms in the
outermost row of the hexagon.  Also of importance are the total
number of sites in the unit cell $N_{\rm Total}$ (equal to the
number of atoms {\it before} the hole is made) and the number of
removed atoms $N_{\rm Removed}$. As an example, for the $\{7,3\}$
lattice $N_{\rm Total}$= 294 and $N_{\rm Removed} = 60$. Below,
results for structures with $L\le 14$  and varying $R$ have been
compiled taking care that no dangling bonds are formed, i.e. that
all atoms have at least two neighbors. While these structures are
too small for present-day lithography, results for realistic
structures are easily obtained by simple scaling laws, as
demonstrated below.

\begin{figure}
\includegraphics[width=0.44\textwidth]{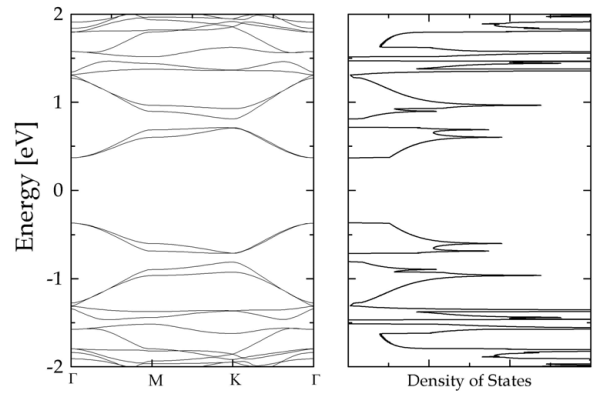}
\caption{Energy band structure and associated density of states for
a $\{7,3\}$ antidot lattice. The notation $\Gamma$, M, and K refers
to high symmetry points of the Brillouin zone.} \label{fig2}
\end{figure}

\begin{figure}
\includegraphics[width=0.44\textwidth]{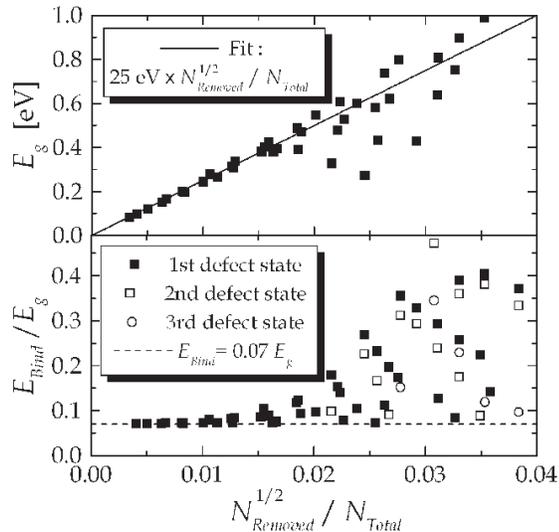}
\caption{Compilation of energy gaps (upper panel) and defect state
binding energies (lower panel). When displayed versus $N^{1/2}_{\rm
Removed}/N_{\rm Total}$, very simple scaling is observed. Note that
$N^{1/2}_{\rm Removed}/N_{\rm Total}$ is small for realistic
structures.} \label{fig3}
\end{figure}

We model the structures using a tight-binding (TB) description
considering a single $\pi$-orbital on each site and assuming a
nearest-neighbor hopping integral of $-\beta$, with $\beta\approx
3.033$ eV~\cite{Sait1998}. In this description, energy levels are
always distributed symmetrically above and below zero, which defines
the Fermi level in the undoped case. The TB approximation is
necessary due to the large antidot cells. It is known to accurately
reproduce the low-energy part of the density-functional (DFT) band
structure of graphene~\cite{Reich2002}. Edges, however, require a
modification of hopping integrals near the edge to ensure agreement
between DFT and TB calculations~\cite{sonprl}. We have checked that
the computed band structures are generally robust against such
modifications, which simply produce a minor additional opening of
the band gap. The electronic band structure and density of states
for the $\{7,3\}$ structure are illustrated in Fig.\ \ref{fig2}.
Importantly, a substantial energy gap of approximately 0.73 eV opens
around the Fermi level~\cite{remark}. Hence, as hinted above, the
periodic perturbation turns the semi-metal into a semiconductor. In
the top panel of Fig.\ \ref{fig3}, band gaps $E_g$ of several
structures are plotted versus the quantity $N^{1/2}_{\rm
Removed}/N_{\rm Total}$. When plotted in this manner, a roughly
linear behavior is observed. This simple result may be rationalized
within the linearized Hamiltonian approximation treating electrons
as massless Dirac fermions subject to the periodic perturbation of
the antidot lattice. In this description, the wave function is a
two-component spinor representing the two sublattices. The
corresponding Hamiltonian is the $2\times 2$ matrix operator
\begin{equation}
H=
\begin{pmatrix}
  V(x,y) & v_F(p_x-ip_y) \\
  v_F(p_x+ip_y) & V(x,y)
\end{pmatrix}
\end{equation}
where $V$ is the periodic antidot potential, $\mathbf{p}$  is the
momentum operator, and the Fermi velocity $v_F=\sqrt{3}\beta
a/(2\hbar)\approx 10^6$m/s. In the absence of a potential, the
energy eigenvalues are simply $E=\pm\hbar v_F|{\mathbf k}|$. If the
potential is approximated by infinite barriers at the positions of
the antidots, the eigenvalue problem is reduced to the form
\begin{equation}
v_F^2(p_x^2+p_y^2)\psi=E^2\psi,
\end{equation}
with the boundary condition that $\psi$  vanishes in the barrier
region. The equation is mathematically similar to the usual
effective mass equation. For an antidot lattice in a usual
semiconductor material such as GaAs, simple scaling arguments lead
to a band gap varying as $E_g\propto A^{-1}_{\rm Total} f(A_{\rm
Removed}/A_{\rm Total})$, where $A_{\rm Total}$ is the area of the
unit cell and $A_{\rm Removed}$ is the area removed inside each unit
cell. In graphene, a similar behavior is expected except that the
linear band structure changes the prefactor from $A^{-1}_{\rm
Total}$ to $A^{-1/2}_{\rm Total}$, i.e.\ $E_g\propto A^{-1/2}_{\rm
Total} g(A_{\rm Removed}/A_{\rm Total})\propto N^{-1/2}_{\rm Total}
g(N_{\rm Removed}/N_{\rm Total})$. The fit in Fig.\ \ref{fig3} shows
that $g$ approximately follows a square root behavior $g(N_{\rm
Removed}/N_{\rm Total})\propto\sqrt{N_{\rm Removed}/N_{\rm Total}}$.
Thus, the net result is a gap varying as $E_g\approx K\times
N^{1/2}_{\rm Removed}/N_{\rm Total}$ with a constant $K\approx 25$
eV. For large unit cells, $N^{1/2}_{\rm Removed}/N_{\rm Total}$ is
small and in this case the linear fit is an excellent approximation.
The weaker scaling ($A^{-1/2}_{\rm Total}$ instead of $A^{-1}_{\rm
Total}$) of graphene is very favorable for the purpose of obtaining
large band gaps even for relatively large structures. The practical
limits of present day e-beam lithography probably restrict the
obtainable size of the unit cell to around 10 nm across
corresponding to a total number of carbon atoms of $N_{\rm
Total}\approx 3000$. Assuming $N_{\rm Removed}\approx N_{\rm
Total}/4$ we find a substantial gap of 0.23 eV. Hence, band gaps
much larger than the thermal energy at room temperature are
certainly realistic. This feature, which is a direct consequence of
the massless Dirac fermion behavior, is very important for the
feasibility of the graphene based devices considered here.

\begin{figure}
\includegraphics[angle=90,width=0.44\textwidth]{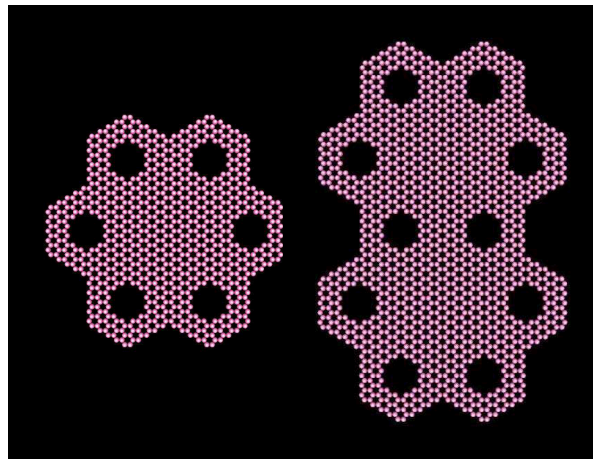}
\caption{Single (left) and double (right) defects for the $\{5,2\}$
antidot lattice. To compute defect states, super cells containing
defects surrounded by six intact units are repeated periodically. }
\label{fig4}
\end{figure}

We now turn to the role of intentional defects in the antidot
lattice produced by leaving one or several unit cells intact, i.e.\
without a hole. Such defects may support localized electronic states
and may consequently be utilized for electron spin qubits, as we
will now demonstrate. An example of single and double defects for
the $\{5,2\}$ structure is shown in Fig.\ \ref{fig4}. For isolated
single defects, we compute localized states by periodically
replicating the super cell consisting of one intact and six
perforated cells illustrated in the figure. The states are
sufficiently localized that cross-talk between neighboring super
cells is negligible. Periodicity is not crucial for the appearance
of bound states \cite{Flin2005}. Defect states are identified by an
energy lying in the fundamental energy gap, i.e.\ the gap containing
the Fermi energy. In fact, other energy gaps may exist as
illustrated in Fig.\ \ref{fig2}; here we focus solely on states in
the fundamental gap. If the gap is sufficiently large (i.e.\ if
$N^{1/2}_{\rm Removed}/N_{\rm Total}$ is large) several defect
states are supported. In the lower panel of Fig. \ref{fig3}, a
compilation of binding energies for the three lowest defect states
is shown. We define the binding energy $E_{\rm Bind}$ as the
downwards shift of the defect state energy measured from the
conduction band edge. Hence, a defect state at the Fermi energy
would have a binding energy of $E_g/2$. For small band gaps, only a
single defect state is supported but several defect states appear in
an irregular pattern as the confinement increases. Note that the
scatter in the data points in the plot reflects actual variations
and not computational inaccuracy. Importantly, the binding energy in
the limit of small band gaps is seen to approach a constant fraction
$\simeq 0.07$ of the energy gap. Hence, for the 10 nm unit cell
considered above, a defect state would be bound by roughly 16 meV.
This implies that liquid nitrogen cooling should be sufficient to
observe these states.

Next, we consider two tunnel coupled defect states in  a ``double
defect'', illustrated in Fig.\ \ref{fig4}. With an electron
occupying a non-degenerate state in each defect, the spins of the
two electrons couple due to the exchange interaction $J {\mathbf
S}_1\!\cdot{\mathbf S}_2$. If the two single-defect states are
energetically aligned, the exchange coupling is given as $J=4t^2/U$
according to the Hubbard approximation. Here, $t$ is the tunnel
coupling between the two defect states, and $U$ is the single-defect
Coulomb energy. As discussed in Ref.\ \cite{Loss1998}, the exchange
coupling constitutes a key element in quantum computing
architectures based on electron spins as qubits, enabling
interactions between different qubits. Importantly, the exchange
coupling can be controlled with external gate potentials. Metallic
gates could be realized by lithographic methods and placed either
below or on top of the graphene sheet but will not be considered
further here. For evaluation of the exchange coupling, we calculate
the single-defect Coulomb energy $U$ by the method presented in
Ref.~\onlinecite{Pede2004} (ignoring overlap between different
atomic $\pi$-orbitals) using the Ohno form to interpolate between
the intra- and long-range inter-atomic Coulomb coupling. A Hubbard
$U_{\pi}$ for carbon $\pi$-orbitals of 20.08 eV~\cite{Pede2004} and
dielectric constant of 2.5~\cite{Ando2006} (as appropriate for
graphene on SiO$_2$) are applied. The tunnel coupling $t$ is
extracted from the single-particle energy spectrum.

\begin{figure}
\includegraphics[width=0.44\textwidth]{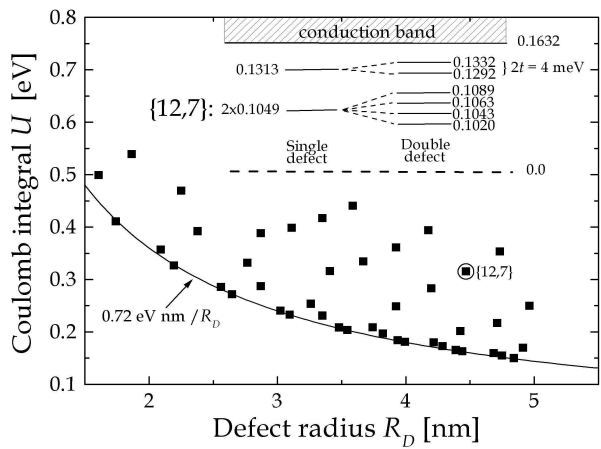}
\caption{Compilation of Coulomb energies $U$ for single defect
states
 showing roughly linear scaling with $R_D^{-1}$ for delocalized states.
Inset: Level structure for single- and double defects in a \{12,7\}
lattice. The Coulomb energy for this case is indicated by the
circle.} \label{fig5}
\end{figure}

Our findings for the Coulomb energy $U$ are illustrated in Fig.\
\ref{fig5}. In the plot, $R_D$ is the effective defect radius
calculated by including half the area of the surrounding cells and
writing the total area as $\pi R_D^2$. The smallest $U$'s are found
for the least localized states for which $U$ scales as the expected
$R_D^{-1}$.  The inset shows, as an example, the single-electron
level diagram for single- and double defects in a $\{12,7\}$
lattice. This structure has $N_{\rm Total}$= 864 and $N_{\rm
Removed} = 348$ and supports two single-defect states. Of these, the
upper one is non-degenerate and the Coulomb energy is 0.315 eV. Due
to the large double defect super cell, this is about the largest
structure that we have been able to analyze. As shown, the level
splitting corresponds to a tunnel coupling of $t\approx 2$ meV
between the two non-degenerate single-defect states. Hence, based on
the $\{12,7\}$ values we may estimate the exchange coupling to be on
the order of $J\approx 50$ $\mu$eV. Naturally, this value could be
tuned by appropriate design of the barrier region that, for
simplicity, has been constructed from two intact unit cells. Also,
going to larger single defects would decrease $U$ and, in turn,
increase $J$. Note, however, that $t$ depends exponentially on
barrier width whereas $U$ is only weakly dependent on geometry.
Hence, the geometric influence on $J$ will be determined mainly
through $t$ rather than $U$.

We believe that the approach outlined above can be extended to more
complicated structures. Going from a single pair of spin qubits in
an isolated double defect to several coupled spins could be achieved
with little added complication. Similarly, a double defect could be
replaced by a linear array of defects. Hence, the number of qubits
can be increased essentially without complicating the fabrication
procedure. In practice, excellent control of the e-beam lithography
process remains a critical issue.

In summary, we have shown that antidot lattices pave the way for
controlled manipulation of the electronic properties of graphene
sheets. The material can be rendered semiconducting with a
significant and controllable energy gap. The magnitude of the gap is
explained by a simple scaling argument and could reach several
tenths of eVs for realistic structures. Introducing defects into the
antidot lattice leads to the formation of localized electronic
states. Combined with the extremely long spin coherence time of
carbon-based materials this could lead to a practical realization of
spin qubits. With a properly designed double defect, two-electron
states derived from defect levels near the Fermi level are found to
fulfil the requirements for such qubits.

\acknowledgements{APJ and CF acknowledge the FiDiPro program of the
Finnish Academy of Sciences for support during the final stages of
this work.  We thank Dr. A. Krasheninnikov for an informative
discussion on gap states in graphene.}


\end{document}